\documentclass[aps,showpacs,nofootinbib,superscriptaddress,preprintnumbers]{revtex4}

\usepackage{graphicx}
\usepackage{amsmath}
\usepackage{bm}
\usepackage{enumerate}
\usepackage{array}

\begin{document}
%\setvruler[11.4pt][1][1][3][1][30pt][30pt][-10pt]

%%%%%%%%%%%%%%%%%%%%%%%%%%%%%%%%%%%%%%%%%%%%%%%%%%%%%%%%%%%%%%%%%%%%%%%%
\title{A method for tuning parameters of Monte Carlo generators and its
  application to the 
  determination of the unintegrated gluon density}
%%%%%%%%%%%%%%%%%%%%%%%%%%%%%%%%%%%%%%%%%%%%%%%%%%%%%%%%%%%%%%%%%%%%%%%%

\author{Alessandro Bacchetta}
\email{alessandro.bacchetta@unipv.it}
\affiliation{Dipartimento di Fisica Nucleare e Teorica, Universit\`a di Pavia,
  and INFN Sezione di Pavia, via Bassi 6, I-27100 Pavia, Italy}

\author{Hannes Jung}
\email{hannes.jung@desy.de}
\affiliation{Deutsches Elektronen-Synchroton DESY, Notkestrasse 85, D-22603
  Hamburg, Germany} 
\affiliation{Departement Fysica, Universiteit Antwerpen (CGB),
  Groenenborgerlaan 171, 
B-2020 Antwerpen}

\author{Albert Knutsson}
\email{albert.knutsson@desy.de}
\affiliation{Deutsches Elektronen-Synchroton DESY, Notkestrasse 85, D-22603
  Hamburg, Germany} 

\author{Krzysztof Kutak} 
\email{krzysztof.kutak@desy.de}
\affiliation{Deutsches Elektronen-Synchroton DESY, Notkestrasse 85, D-22603
  Hamburg, Germany} 
\affiliation{Departement Fysica, Universiteit Antwerpen (CGB),
  Groenenborgerlaan 171, 
B-2020 Antwerpen}

\author{Federico von Samson-Himmelstjerna}
\email{federico.samson-himmelstjerna@desy.de}
\affiliation{Deutsches Elektronen-Synchroton DESY, Notkestrasse 85, D-22603 Hamburg, Germany}

%%%%%%%%%%%%%%%%%%%%%%%%%%%%%%%%%%%%%%%%%%%%%%%%%%%%%%%%%%%%
\begin{abstract}

A method for tuning parameters in Monte Carlo generators is described and
applied to a specific case.
The method works in the following way: each observable is generated several
times using different values of the parameters to be tuned. 
The output is then approximated by some analytic form to describe 
the dependence of the observables on the parameters. 
This approximation is used  
to find the values of the parameter that give the best description of the
experimental data. 
This
results in significantly faster fitting compared to an approach 
in which the generator is called iteratively.
As an application, we employ this method to
fit the parameters of the unintegrated gluon density used 
in the {\sc Cascade} Monte
Carlo generator, using inclusive deep inelastic 
data measured by the H1 Collaboration.
We discuss the results of the fit, its limitations, and its strong points.
\end{abstract}

%%%%%%%%%%%%%%%%%%%%%%%%%%%%%%%%%%%%%%%%%%%%%%%%%%%%%%%%%%%%%%%%%%%%%%%%


%\date{\today}

\preprint{DESY 10-013}
%Version of \today, \currenttime}

\maketitle

%%%%%%%%%%%%%%%%%%%%%%%%%%%%%%%%%%%%%%%%%%%%%%%%%%%%%%%%%%%%
\section{Introduction}
\label{s:intro}

The substructure of the proton is parametrized in terms of parton
distribution functions (PDFs). In perturbative QCD 
the PDFs are
given by solutions of integral equations, for which the
initial input distributions have to be determined by global fits to the
available experimental data (see, e.g., \cite{Vogt:2007vv} and references
therein). 
All present global fits are based on fixed-order
calculations in $\alpha_s$, the strong coupling constant, and on factorization
theorems that apply to specific inclusive processes, where most of the
final-state properties are integrated over.

To study more exclusive processes (i.e., multiparticle production or
multidifferential cross sections), Monte Carlo (MC) event generators are
used. The physics included in the generators is often not the same as the one
described by the factorization theorems and used in global fits. 
For instance, most of the generators
do not implement complete 
next-to-leading-order (NLO) QCD corrections, but on the other hand they
implement parton showers, which partially take into account all-orders
resummation effects. 

Due to these differences, in principle using in the MC generators 
the PDFs extracted from global fits is not fully consistent. Ideally, the
PDFs should be fitted directly using a MC event
generator~\cite{Jung:2008}, together with all other extra 
parameters of the generator. 
Unfortunately, the parameters of a generator are difficult to tune 
efficiently because
minimization programs require several sequential calls of the generator. This
can be extremely time-consuming, especially for more
exclusive events.

Motivated by Refs.~\cite{Abreu:1996na}, we are using a 
fast and  efficient method to fit generator parameters. 
The method is
based on using a MC event generator to produce a
grid in parameter space for each observable. The parameter
dependence is then approximated by polynomials
before the fit is performed, which significantly reduces the fitting
time. This method has also been recently used in Ref.~\cite{Buckley:2009bj}.

As an application, we tune the parameters of the unintegrated gluon
distribution function (also called transverse-momentum-dependent gluon distribution
function) using the {\sc Cascade} MC event
generator~\cite{Jung:2001hx}, 
by fitting the generator predictions to inclusive deep inelastic scattering
data measured by the H1 Collaboration~\cite{Adloff:2000qk}. 
We explore the reliability and the limitations of the method and
study to which extent the data can constrain the input parameters. 

The paper is organized as follows.
In section~\ref{Section: Method} we give the details of the fitting method.
We give a simple example which we generalize to several parameters and
observables. In section~\ref{Section: CASCADE}, we discuss how the fitting
method is applied to a specific case
and the results of the
tune are presented in section~\ref{Section: Results}. We draw some
conclusions at the end of the paper.

%%%%%%%%%%%%%%%%%%%%%%%%%%%%%%%%%%%%%%%%%%%%%%%%%%%%%%%%%%%%%%%%%%%%%%%%%%%
\section{Tuning method}
\label{Section: Method}
%%%%%%%%%%%%%%%%%%%%%%%%%%%%%%%%%%%%%%%%%%%%%%%%%%%%%%%%%%%%%%%%%%%%%%%%%%%

In general, the goal of the tuning is to describe a set of $N$ experimental 
observables
$Y_i^{\rm{ex}}$, 
with errors $\delta Y_i^{\rm{ex}}$, by means
of a theoretical model 
(in this case a MC generator) that depends on the parameters
$\alpha_a$, 
and predicts the observables to be
 $Y_i^{\rm{MC}}(\alpha_a)$, 
with errors $\delta Y_i^{\rm{MC}}(\alpha_a)$. 
The values of the parameters that give the best
description of the data can be found by minimizing the $\chi^2$ function
\begin{equation} 
\label{e:chi2brute}
\chi^2(\alpha_a)=\sum_{i=1}^N
\frac{\bigl[Y_i^{\rm{MC}}(\alpha_a)
                  -Y_i^{\rm{ex}}\bigr]^2}
{\bigl[\delta Y_i^{\rm{MC}}(\alpha_a)\bigr]^2
                  +\bigl[\delta Y_i^{\rm{ex}}\bigr]^2}.
\end{equation} 
Usually, the minimization is done by numerical programs such as
MINUIT~\cite{James:1975dr}. The generator predictions have to be
computed typically a few hundred times for different choices of the parameters
before the minimum is found. This ``brute-force'' procedure is highly time
consuming.  

An alternative approach has been used in, e.g.,
Ref.~\cite{Braunschweig:1988qm} as early as twenty years ago, and more
recently in Refs.~\cite{Abreu:1996na,Buckley:2009bj}. 
First, for each observable a grid in
parameter space is built, running the MC generator with several
values of the parameters. Secondly, the grids are approximated by analytic
functions of the parameters, usually polynomials. 
These functions give a fair description of
the generator output and can be used in its stead. In this way, 
finding the
parameter values that best fit the data becomes a much faster task.

The method turns out to be particularly time efficient. A fitting procedure
typically requires to sequentially 
calculate $\chi^2$ a few hundred times for different
values of the parameters.  Building the grids in parameter space also requires
running the MC generator a few hundred times, but each computation can be done
independently in parallel. Once the grid is built and approximated
analytically, minimizing the $\chi^2$ is extremely fast. It becomes very
convenient to run the minimization with different initial values of the
parameters, or including only a subsample of the observables. However, if new
data points are added, a new grid has to be produced for each new data
point.

%%%%%%%%%%%%%%%%%%%%%%%%%%%%%%%%
\subsection{A simple example}

\begin{figure}
%\centerline{
\includegraphics[width=0.7\columnwidth]{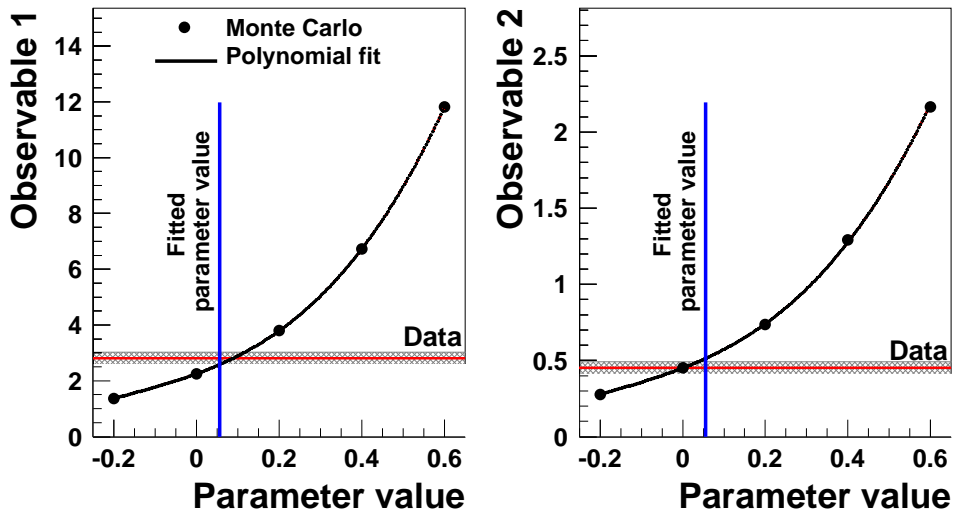}
%}	   
\caption{
Example of the fit procedure applied to a single parameter and two
observables: the horizontal lines with bands represent the
experimental values and errors of the observables, 
the points indicate the grids predicted by the 
MC generator for different values of the parameter on the $x$ axis, 
the curved lines represent analytical approximations to the
grids, and the vertical lines indicate the best-fit value of the parameter.
}
\label{Fig:example}
\end{figure}

To illustrate the method, we start from a simple example. Suppose 
we need to fit two
data points $Y_1^{\rm{ex}}$ and $Y_2^{\rm{ex}}$ with their errors
(e.g., two cross-section measurements) using a
MC generator with one tunable parameter $\alpha$. In Fig.~\ref{Fig:example}, we
indicate the two data points with solid horizontal lines with their
error bands.

First, we choose 
5 values ($j=1,\ldots,5$) of the parameter $\alpha$ and
generate 5 predictions for each observable, i.e.,
two grids
$\bigl(\alpha_{1,j},Y_1^{\rm{MC}}(\alpha_{1,j})\bigr)$ and 
$\bigl(\alpha_{1,j},Y_2^{\rm{MC}}(\alpha_{1,j})\bigr)$, with
statistical errors due to the Monte Carlo method. In
Fig.~\ref{Fig:example}, these grids are indicated by points (the errors are too
small to be visible).

 Then we choose an analytical form to approximate the two grids, which
 will be a function of $\alpha$, but also of two new sets of parameters
 $A_1,B_1,\ldots$ and $A_2,B_2,\ldots$, one for each grid.  To avoid
 confusion, with denote these new parameters as ``grid parameters,'' to be
 distinguished from the original MC parameters. 
In principle, the functional form itself could be different for each distinct
grid, but in practice it is more convenient to choose the same form. 
 To make the procedure easier, 
it is a good idea to choose a function that is linear in the
 grid parameters, for instance a third-degree polynomial
\begin{equation}
Y_i^{\rm{app}}(\alpha;A_i,B_i,C_i,D_i) = A_i + B_i\, \alpha + C_i\, \alpha^2 
+ D_i \,\alpha^3 .
\label{e:polyexample}
\end{equation} 

The best values
of the grid parameters are chosen by means of a 
 $\chi^2$ minimization for each separate grid. We define
 this procedure as ``grid approximation,'' to be distinguished from the actual
 fit to the experimental data. 
We define in this case
\begin{equation} 
\chi^2_i(A_i,B_i,\ldots)=\sum_{j=1}^5
 \frac{\bigl[Y_i^{\rm{app}}(\alpha_{1,j};A_i,B_i,\ldots)
                           -Y_i^{\rm{MC}}(\alpha_{1,j})\bigr]^2} 
 {\bigl[\delta Y_i^{\rm{MC}} (\alpha_{1,j}) \bigr]^2}. 
\end{equation} 
The polynomials obtained using the best-fit parameter values,
$\hat{A}_i, \hat{B}_i, \ldots$ are indicated as a curved solid line in
Fig.~\ref{Fig:example}. The $\chi_i^2$ analysis allows us also to estimate the
errors bands on the grid approximations (not visible in the
figure).  
 
At this point, it is useful to remark that 
the degree of the polynomial introduced in  Eq.~\ref{e:polyexample} is a
matter of choice. Usually, the higher the degree, the better the description of
the grids
becomes. However, from a certain point on, adding an extra degree does not
improve 
the quality of the approximation significantly, i.e., it does not change
significantly the sum of the minimum $\chi_i^2$. 
In the case shown in Fig.~\ref{Fig:example}, 
it turns out that a third-degree polynomial gives a much better
description of the grid than a second-degree polynomial, while the 
fourth-degree polynomial does not significantly improve the situation.

Once we have analytical approximations of the Monte Carlo
generated grids, we can finally fix the best value of the
parameter $\alpha$ by
minimizing the function
\begin{equation} 
\chi^2(\alpha)=\sum_{i=1}^2
\frac{\bigl[Y_i^{\rm{app}}(\alpha;\hat{A_i},\hat{B}_i,\ldots)
                  -Y_i^{\rm{ex}}\bigr]^2}
{\bigl[\delta Y_i^{\rm{app}}(\alpha;\hat{A_i},\hat{B}_i,\ldots)\bigr]^2
                  +\bigl[\delta Y_i^{\rm{ex}}\bigr]^2}.
\end{equation}  
In Fig.~\ref{Fig:example} the best-fit parameter value, $\hat{\alpha}_1$, is
indicated as a straight vertical line.

%%%%%%%%%%%%%%%%%%%%%%%%%%%%%%%%%%%%%
\subsection{The general case}

Generalizing the above example, with $N$ experimental points 
(denoted by the index $i$) and $P$ parameters (denoted by the index $a$),
we need to build $N$ grids in
$(P+1)$-dimensional spaces, 
$\bigl(\alpha_{a,j_a},Y_i^{\rm{MC}}(\alpha_{a,j_a})\bigr)$. 
If we choose $J_a$
points for each parameter, the generation of the grid requires 
$J=\prod_{a=1}^PJ_a$  Monte Carlo runs.
Once the grids are built, we approximate 
them using polynomials of degree $n$ (for simplicity we show here explicitly
only the terms up to second degree)
\begin{equation} 
Y_i(A_i,B_{i,a},C_{i,ab},\ldots)=A_i+\sum_{a=1}^P B_{i,a}\,\alpha_a
+\sum_{a=b}^{P}\sum_{b=1}^PC_{i,ab}\,\alpha_a \alpha_b+\ldots \,.
\end{equation} 
Note that the Monte Carlo parameters $\alpha_a$ are the variables of the
polynomials, while the grid parameters are the coefficients.
For degree two and higher, the off-diagonal terms like $C_{i,ab}$, $a\ne b$, 
take into account correlations between the Monte Carlo parameters. 
In our application, we found that third-degree polynomials give a good
description of the grid. Advancing to fourth-degree polynomials does not lead
to significant improvements. 

The total number of coefficients for a degree-$n$ polynomial of
$P$ parameters is $M=\sum_{k=0}^n
\frac{(P+k)!}{k!P!}$. For instance, a polynomial of third degree of four
Monte Carlo parameters has 35 coefficients. 
For simplicity, we denote them
collectively as $A_{i,s}$, where $A_{i,1}=A_i$, $A_{i,s}=B_{i,a}$ for
$s=2,\ldots,P+1$, $A_{i,s}=C_{i,ab}$ for $s=P+2,\ldots,P+2+P(P+1)/2$, etc.

The values of the coefficients that give the best approximation to the grid
are obtained by minimizing
\begin{equation} 
\label{e:chi2approx}
\chi^2_i(A_{i,s})=\sum_{j=1}^J
 \frac{\bigl[Y_i^{\rm{app}}(\alpha_{k,j};A_{i,s})
                           -Y_i^{\rm{MC}}(\alpha_{k,j})\bigr]^2} 
 {\bigl[\delta Y_i^{\rm{MC}} (\alpha_{k,j}) \bigr]^2}. 
\end{equation} 
Since the fit function is linear in the coefficients, the best way to perform
the $\chi^2$ minimization is to use Singular Value
Decomposition (SVD)~\cite{Press:1992}. 
SVD is based on the fact that the relation between the observables
and the grid parameters $A_{i,s}$ can be written as an
over-determined system of linear equations. SVD provides a solution to this
system in a least-squares sense. Compared to other, more general
numerical minimization procedures (such as the ones implemented in MINUIT),
SVD is faster and 
guarantees that the true $\chi^2$ minimum is found. The solution does
not depend on the choice of the initial values of the parameters. This is
particularly important when the minimization involves several dozens of
parameters. 

%Typically, the $\chi_i^2$ minima are very large, due to the fact that the 
%generated points have very small statistical errors. It is difficult then to
%assess a priori if the grids are approximated sufficiently well for the
%purpose of fitting the experimental data. A good test is to check,
%at the end of the entire fitting procedure, if the results obtained using the
%approximated grids can be reproduced by the Monte Carlo directly.

The approximation procedure returns the best-fit values, $\hat{A}_{i,s}$, of
the coefficients and a covariance matrix that can be used to estimate the
statistical error bands on the approximation, $\delta
Y_i^{\rm{app}}(\alpha_a;\hat{A}_{i,s})$  by means of error propagation. 

Once the grids are approximated by polynomials in Monte Carlo parameter space, 
we finally want to choose the values of the parameters $\alpha_a$
that give the best description
of the data. To correctly take into
account systematic errors in the experimental measurements, the $\chi^2$
function has been computed using~\cite{Stump:2001gu}
\begin{equation} 
\chi^2=
\sum_{i=1}^N
\frac{\bigl[Y_i^{\rm{app}}(\alpha_a;\hat{A}_{i,s})
                  -Y_i^{\rm{ex}}+\sum_{k=1}^{n_{{\rm sys}}} {r'_k}^2 \bigr]^2}
{\bigl[\delta Y_i^{\rm{app}}(\alpha_a;\hat{A}_{i,s})\bigr]^2
                  +\bigl[\delta Y_i^{\rm{ex}}\bigr]^2}
+\sum_{k=1}^{n_{{\rm sys}}} {r'_k}^2 ,
\label{e:chi2exp}
\end{equation} 
where the random parameters $r'_k$ are defined in App.~A.
The minimization is done in this case using MINUIT,
since the dependence on parameters $\alpha_a$ is
non-linear. 

The tuning method studied in Ref.~\cite{Buckley:2009bj} is essentially the
same as the one considered here.
The main differences between the two implementations 
reside in the definition of the $\chi^2$
function, which in our case include the statistical error in the grid
approximation ($\delta Y_i^{\rm{app}}$) and the contribution of correlated
systematic uncertainties.

%%%%%%%%%%%%%%%%%%%%%%%%%%%%%%%%%%%%%%%%%%%%%%%%%%%%%%%%%%%%%%%%%%%%%%%%%%%
\section{An application: fitting the unintegrated gluon distribution function
  in {\sc Cascade}} 
\label{Section: CASCADE} 
%%%%%%%%%%%%%%%%%%%%%%%%%%%%%%%%%%%%%%%%%%%%%%%%%%%%%%%%%%%%%%%%%%%%%%%%%%%

The fitting method described before is general and may be
applied to tune any parameter in any Monte Carlo generator. 
At present, however, we want to concentrate on tuning the parameters of the
unintegrated gluon distribution function (uGDF) -- also known as
transverse-momentum-dependent gluon distribution function --  
in the {\sc Cascade} MC generator.

A brief introduction to the  {\sc Cascade} event generator is in order.
For a more
detailed description we refer the
reader to \cite{Jung:2001hx}. 
{\sc Cascade} is a hadron level Monte Carlo event generator for $e p$, $\gamma p$ and $p\overline p$ processes, which
uses the CCFM evolution equation for the initial state parton shower
supplemented with off-shell matrix elements for the hard scattering. To
simulate the hadronization process, {\sc Cascade} uses the Lund
string model~\cite{Andersson:1983ia}.

The CCFM equation is a linear
integral equation which sums up the cascade of gluons under
the condition that subsequent emissions are angularly
ordered. 
With this ordering it interpolates between DGLAP (resummation of transverse momenta $\alpha_s^n \ln^nk_t^2$) 
and BFKL (resummation of longitudinal momenta $\alpha_s^n \ln^n x$) limits.

\begin{figure}[t!]
\centerline{
\includegraphics[width=0.4\columnwidth]{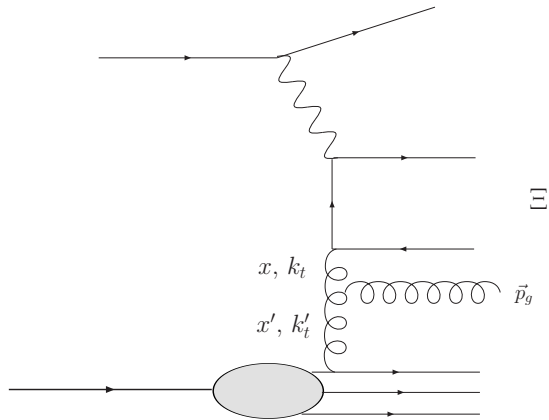}
}	   
\caption{Schematic view of a parton ladder illustrating the
kinematic variables used in the text.}\label{Fig: CCFM}
\end{figure}

In Fig.~\ref{Fig: CCFM} we show schematically a parton ladder defining the 
kinematic variables which we use in equations
below.
The CCFM equation reads:
\begin{equation} 
A(x,k_t,\overline{q})=A_0(x,k_t,\overline{q})+\int_x^1\frac{dz}{z}\int\frac{d^2q}{\pi q^2}\Theta(\overline{q}-z q)\Delta_s(\overline{q},z q)
\tilde P_{gg}(z,q,k_t)A\bigg(\frac{x}{z},k_t',q\bigg)
\label{e:CCFM}
\end{equation}
where $A_0(x,k_t,\overline{q})$ is the input distribution, $x$
denotes the longitudinal momentum fraction of the proton
carried by the gluon, $k_t$ is the 2-dimensional transverse
momentum of the $t$ channel gluon, $z=x/x'$ is the splitting
variable, $\overline{q}$ is the factorization scale specified by the
maximum allowed angle $\Xi$ between the partons in the matrix elements, 
$ k_t'=| \vec k_t +(1-z) \vec
q|$.  We also introduced $q$ as a shorthand notation for the
2-dimensional momentum $\vec q\equiv \vec q_t=\vec
p_g/(1-z)$. The Sudakov form factor (which we do not write
explicitly) $\Delta_s(\overline{q}, z q)$ for inclusive quantities
regularizes the $1/(1-z)$ collinear singularity of the
splitting function  $\tilde P_{gg}(z,q,k_t)$.
%
%For a description of the implementation of the CCFM equations in {\sc Cascade} 
%we refer the
%reader to \cite{Jung:2001hx}. 

The input distribution can be written as
\begin{equation} 
A_0(x,k_t,\overline{q}) = A_0(x,k_t)\,\Delta_s\bigl(\overline{q},Q_0\bigr).
\end{equation} 
We choose to parametrize the distribution at the
starting scale $Q_0 = 1.2$ GeV in the following way
\begin{equation}
x A_0(x,k_t)=N x^{-B} (1-x)^C (1-D x) e^{-(k_t-\mu)^2/\sigma^2}
\label{Eq: startdist}
\end{equation}
where $N,B, C, D, \mu,\sigma$ should be in principle determined from fits. In
practice, for the purpose of the present study we fix
$C=4$, $\mu=0$ GeV, $\sigma=1$ GeV~\cite{Jung:2001hx}. 
The value of parameter $C$ is dictated by the spectator counting
rules\cite{Brodsky:1973kr}: 
since at large $x$ gluons are suppressed as compared to
quarks, $C$ for gluons has to be larger than 3. Previous studies suggests
$C=4$~\cite{Jung:2006ji}. 
Parameter $D$, typically included in global fits of the PDFs (see, e.g.,
\cite{Gluck:2007ck,Martin:2009iq}), was set to zero in earlier studies with
{\sc Cascade}~\cite{Hansson:2003xz,Jung:2004gs,Jung:2006ji}.  As we
will show later, the addition of this parameter substantially
improves the description of the data we consider. 

%%%%%%%%%%%%%%%%%%%%%%%%%%%%%%%%%%%%%%%%%%%%%%%%%%%%%%%%%%%%%%%%%%%%%%%%%%%
%\section{Fitting the unintegrated gluon distribution function to data}
%\label{Section: Fitting}
%%%%%%%%%%%%%%%%%%%%%%%%%%%%%%%%%%%%%%%%%%%%%%%%%%%%%%%%%%%%%%%%%%%%%%%%%%%

The parameters of the starting uGDF, $N$, $B$, and
$D$ in Eq.~\ref{Eq: startdist}, are determined by
fits to the $F_2$ structure function in inclusive deep inelastic
scattering,  $ep \to e' X$, as measured by the H1
Collaboration~\cite{Adloff:2000qk}. We chose this data set in order to compare
our results with earlier determinations of the uGDF. 
The measurement was made 
at the electron-proton center of mass energy
$\sqrt{s}=300.9$~GeV within the kinematic range $1.5 < Q^2 <
150$~GeV$^2$, $3 \times 10^{-5}<x_{Bj}<0.2$. Here
$Q^2$ is the virtuality of the
exchanged boson, and $x_{Bj}$ is the Bjorken scaling
variable. 
The measurements cover the small-$x_{Bj}$ region where gluon-induced 
processes
dominate and we should have a good sensitivity to the values
of the parameters in the uGDF. In total, there are 122 data points binned in
$x_{Bj}$ and $Q^2$. 

We considered two different cases: in the first case we
restricted ourselves to $x_{Bj}\leq 0.005$ and $Q^2 \geq 4.5$ GeV$^2$, as in
most of the available {\sc Cascade} tunes~\cite{Hansson:2003xz,Jung:2004gs,Jung:2006ji}; 
%in order 
%reproduce the results obtained in Ref.~\cite{Jung:2006ji}, 
%where the MC generator was
%used directly; 
in the second case, we extended the range to the whole data set.

In summary, we performed four kinds of fits: 
\begin{enumerate}[{Fit} 1. ]
\item{$x_{Bj}\leq 0.005$ and $Q^2  \geq 4.5$ GeV$^2$, $D=0$ in Eq.~\eqref{Eq:
      startdist},}
\item{$x_{Bj}\leq 0.005$ and $Q^2  \geq 4.5$ GeV$^2$, $D\neq0$ in Eq.~\eqref{Eq:
      startdist},}
\item{full $x_{Bj}$ and $Q^2$ range, $D=0$ in Eq.~\eqref{Eq:
      startdist},}
\item{full $x_{Bj}$ and $Q^2$ range, $D \neq 0$ in Eq.~\eqref{Eq:
      startdist}.}
\end{enumerate}

The grid in parameter space was built in two different ways depending on
whether parameter $D$ was set to zero or treated as a fit parameter. The final
grids were chosen after performing rough fits with wider grids.
In Fit 1 and 3, we chose
$N=[0.5,0.55,0.6,0.65,0.7,0.75,0.8,0.85,0.9]$ and
$B=[-0.05,-0.025,0.0,0.025,0.05,0.075,0.1,0.125]$
for a total of 72 grid
points. 
In Fit 2 and 4, we
chose 
$N=[0.30,\ 0.38,\ 0.46,\ 0.54,\ 0.62,\ 0.70]$,
$B=[0.00,\ 0.05,\ 0.10,\ 0.15,\ 0.20]$, and 
$D=[-12, -10 ,-8, -6, -4 ,-2, 0]$, 
for a total of 150 grid points.
For each point 2.5 million events were generated. Each generation takes
a few hours of computing time and can be run in
parallel. 

Describing the grid with a third degree polynomial is in our experience the
best choice. The quality of the grid approximation is very good, with an
average $\chi^2/{\rm n.d.f}$  of 
1.08, 1.05, 1.11, 1.12 
for Fit 1 to 4,
respectively. We studied the performance of polynomials of different degree. 
At variance with
Ref.~\cite{Buckley:2009bj}, we observed that second-degree polynomials do not give a
sufficiently good description of the grid. 
Fourth-degree polynomials perform better but do not lead to a significant
improvement of $\chi^2$. 

Using the covariance matrix obtained in the approximation procedure,
the errors of the coefficients in the polynomials are propagated as
theoretical errors to the observables we need to fit, denoted as $\delta
Y^{\rm app}$ in Eq.~\eqref{e:chi2exp}.

Once the parameter dependence was described by the
polynomials, the parameters were fitted to the data by using
MINUIT, using the Migrad method~\cite{James:1975dr}. 
Approximately 150 iterations were needed by the
program in order to find the lowest $\chi^2$ within the
allowed limits set by the grid. This minimization took only few seconds.

If our analytical grid description is good enough, we can expect the number of
iterations to be the same to if we used the true MC instead of the
polynomial approximation.
 Assuming that running the generator once with
the current statistics takes approximately 6 hours of CPU
time, fitting the Monte Carlo parameters with a
conventional iterative way is expected to take 150$\times$6
hours. Clearly, in such case one is forced to
drastically reduce the statistics, and the fit could be influenced by
statistical fluctuations. In addition, our method allowed us
to quickly remake the fit by feeding MINUIT with different starting
values. In this way we reduced the risk of finding a local
minimum.

%%%%%%%%%%%%%%%%%%%%%%%%%%%%%%%%%%%%%%%%%%%%%%%%%%%%%%%%%%%%%%%%%%%%%%%%%%%%%%%%%%%%%%%%%%%%%%%%%%%%%%%%%%%%%%%%%%%%%%%%%%%%%%%%%%%%%%%%%%%%%%%%%%%%%%%
\section{Results}
\label{Section: Results}
%%%%%%%%%%%%%%%%%%%%%%%%%%%%%%%%%%%%%%%%%%%%%%%%%%%%%%%%%%%%%%%%%%%%%%%%%%%%%%%%%%%%%%%%%%%%%%%%%%%%%%%%%%%%%%%%%%%%%%%%%%%%%%%%%%%%%%%%%%%%%%%%%%%%%%%

The best-fit values of the parameters are quoted in 
Tab.~\ref{t:fitresults}.  

\setlength{\tabcolsep}{10pt}
\renewcommand{\arraystretch}{2}
\begin{table}[h]
\begin{tabular}{|r|c|c|c|c|c|}
\hline
   & Range & $N$  (GeV$^{-2}$)  &  $B$  &  $D$  &  $\chi^2/{\rm n.d.f.}$
\\ \hline
Fit 1 &\parbox[m][28pt][c]{2.4cm}{$x_{Bj}\leq 0.005$, $Q^2  \geq 4.5$ GeV$^2$}
     & $0.805 \pm 0.032$   & $0.030 \pm 0.006$  & 0 (fixed) & 2.0
\\ \hline
Fit 2 &\parbox[m][28pt][c]{2.4cm}{$x_{Bj}\leq 0.005$, $Q^2  \geq 4.5$ GeV$^2$}
     & $0.417 \pm 0.030$   & $0.125 \pm 0.010$  & $-9.2 \pm 1.3$  & 1.6
\\ \hline
Fit 3 & full
     & $0.582 \pm 0.016$  & $0.070 \pm 0.004$  & 0 (fixed) & 6.2
\\ \hline
Fit 4 & full
     & $0.368 \pm 0.015$  & $0.140 \pm 0.006$  & $-8.03 \pm 0.66$ & 4.6
\\ \hline
\end{tabular}
\caption{Best fit parameters and $\chi^2/{\rm n.d.f.}$ for the fits described
  in the text. 
%The computation of the $\chi^2/{\rm n.d.f.}$ was performed
%using directly the Monte Carlo generator with our best-fit parameters and not
%the grid approximation. 
}
\label{t:fitresults}
\end{table}

\begin{figure}[h]
\centerline{
\includegraphics[width=0.8\columnwidth]{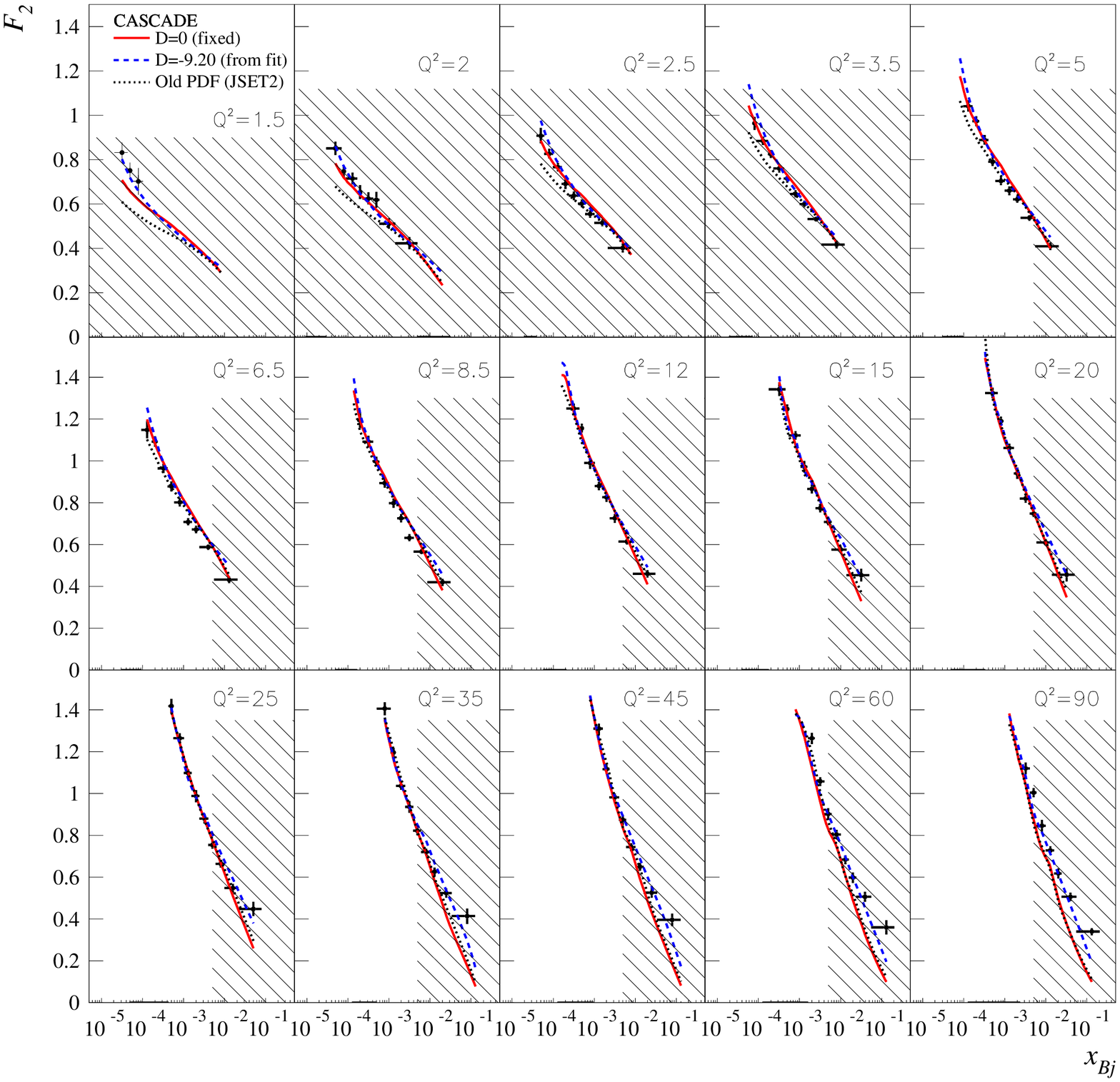}
}	   
\caption{$F_2(x)$ structure function measured by the H1
  Collaboration~\cite{Adloff:2000qk} 
  together with simulations based on the {\sc Cascade} event
  generator, using the unintegrated gluon PDF obtained in
  Ref.~\cite{Jung:2006ji} (dashed line), 
  and using the parameters obtained in Fit 1
  of the 
  present work (solid line). The hatched areas were excluded from the fit.
}
\label{f:case1}
\end{figure}

\begin{figure}[h]
\centerline{
\includegraphics[width=0.8\columnwidth]{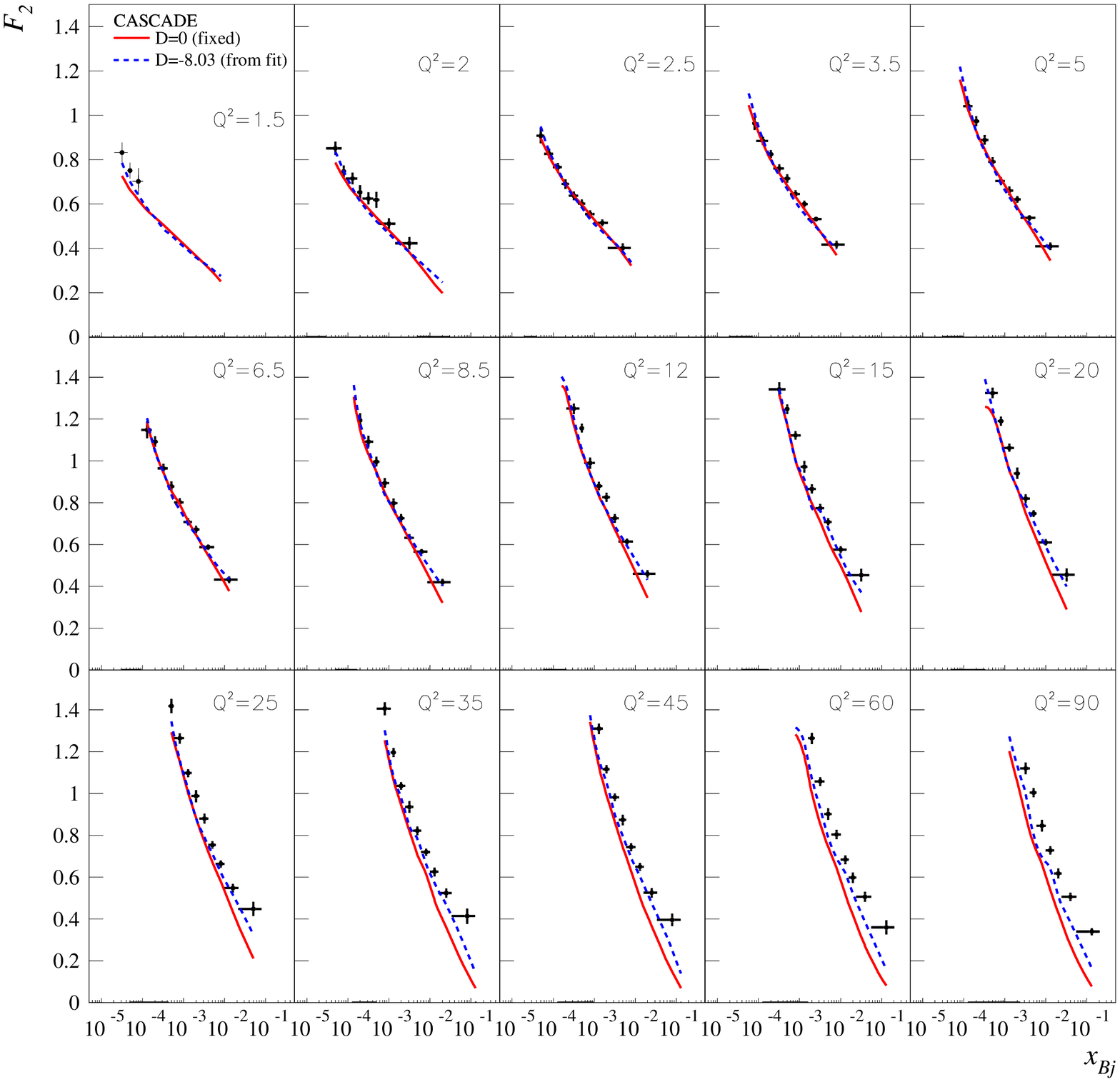}
}	   
\caption{$F_2(x)$ structure function measured by the H1
  Collaboration~\cite{Adloff:2000qk} 
  together with simulations based on the {\sc Cascade} event
  generator, using the parameters of the Fit 3 (dashed line), 
  and Fit 4 (solid line)
  of the 
  present work. In contrast to Fig.~\ref{f:case1}, the whole $x_{Bj}$ and $Q^2$ range has been
  included in the fit.
}
\label{f:case3}
\end{figure}

To have an idea of the performance of our fit, we can compare Fit 1 with
earlier uGDF fits~\cite{Hansson:2003xz,Jung:2004gs,Jung:2006ji}. In
particular, we chose to compare our fit to the J2003 set 2 (JSET2) uGDF~\cite{Hansson:2003xz},
which is the one with the closest conditions to ours. In that set,
parameter $B$ is set to zero. To compare the quality of the description, we ran {\sc Cascade} with a
statistics of 2.5M events, and we found that
our fit gives a $\chi^2/{\rm n.d.f.}=1.4$, while the old set gives
$\chi^2/{\rm n.d.f.}=2.1$. 
In other words, we found
parameters that give a 
better description of the data,
giving us confidence in our fitting method.
% 
%The values of the parameters in Fit 1 are close the the ones obtained in
%Ref.~\cite{Jung:2006ji}, where $B$ and $D$ where set to 0
%the brute-force method was used with $D=0$ and with limited MC statistics, giving 
%$\chi^2/{\rm n.d.f.}= 1.83$. By using
%even higher statistics  we
%obtain a better $\chi^2/{\rm n.d.f.}=1.2$. In other words, we
%were able to 
In
Fig.~\ref{f:case1} we show the results of Fit 1, Fit 2, and JSET2 
 compared to the data.
The results of Fit 3 and 4 are shown in Fig.~\ref{f:case3}.

\begin{figure}[h]
\centerline{
\includegraphics[width=0.33\columnwidth]{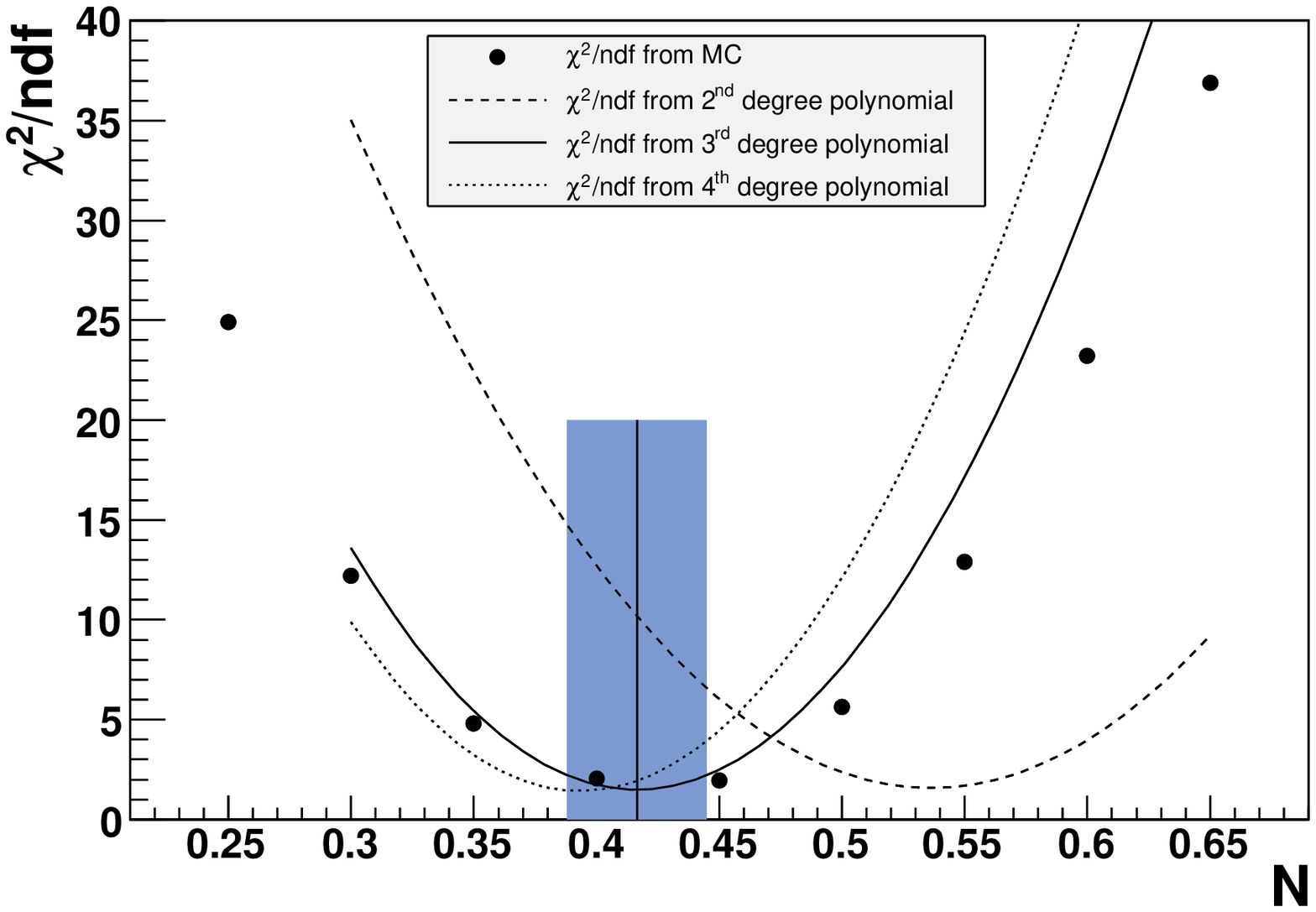}
\hfill
\includegraphics[width=0.33\columnwidth]{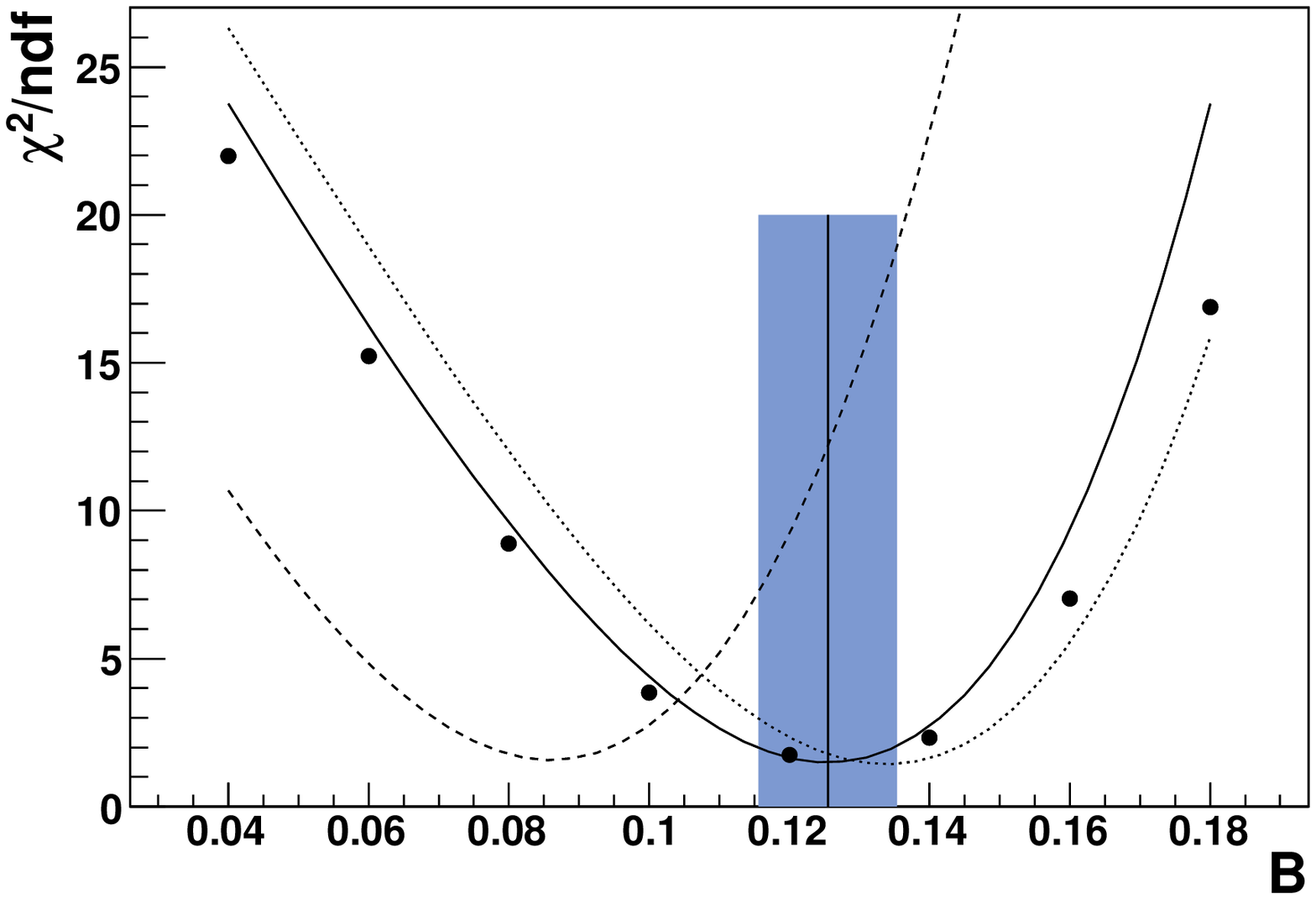}
\hfill
\includegraphics[width=0.33\columnwidth]{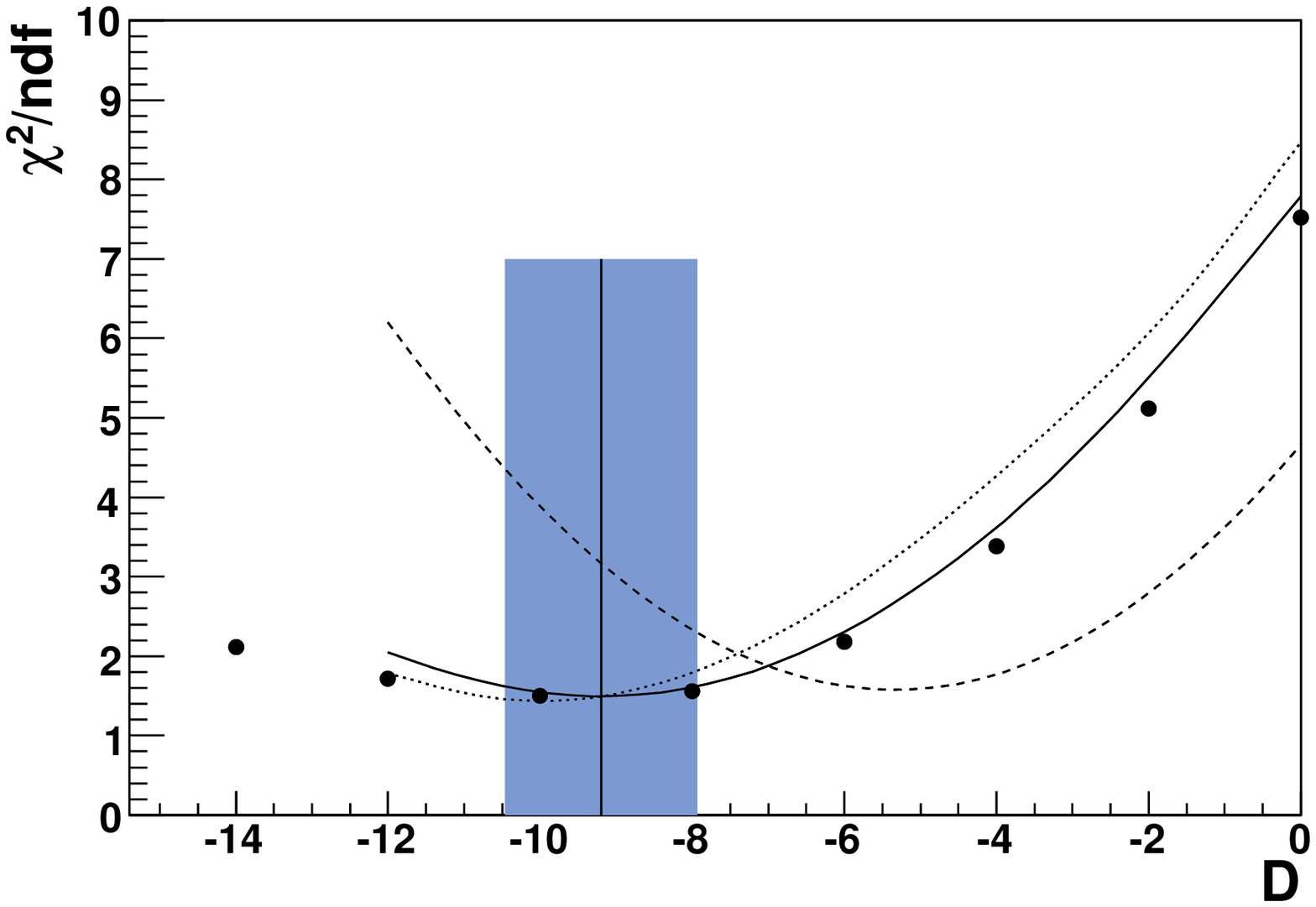}
}
\caption{$\chi^2$ profiles as a function of the parameters of the input
  uGDF for Fit 2. Dots: using the MC generator directly. Lines: using three different
  versions of the polynomial approximation. The vertical line and band
  indicates the position of the minimum and its error (obtained using the
  third-degree polynomial approximation).}
\label{f:chi2scans}
\end{figure}

\begin{figure}[h]
\centerline{
\includegraphics[width=0.33\columnwidth]{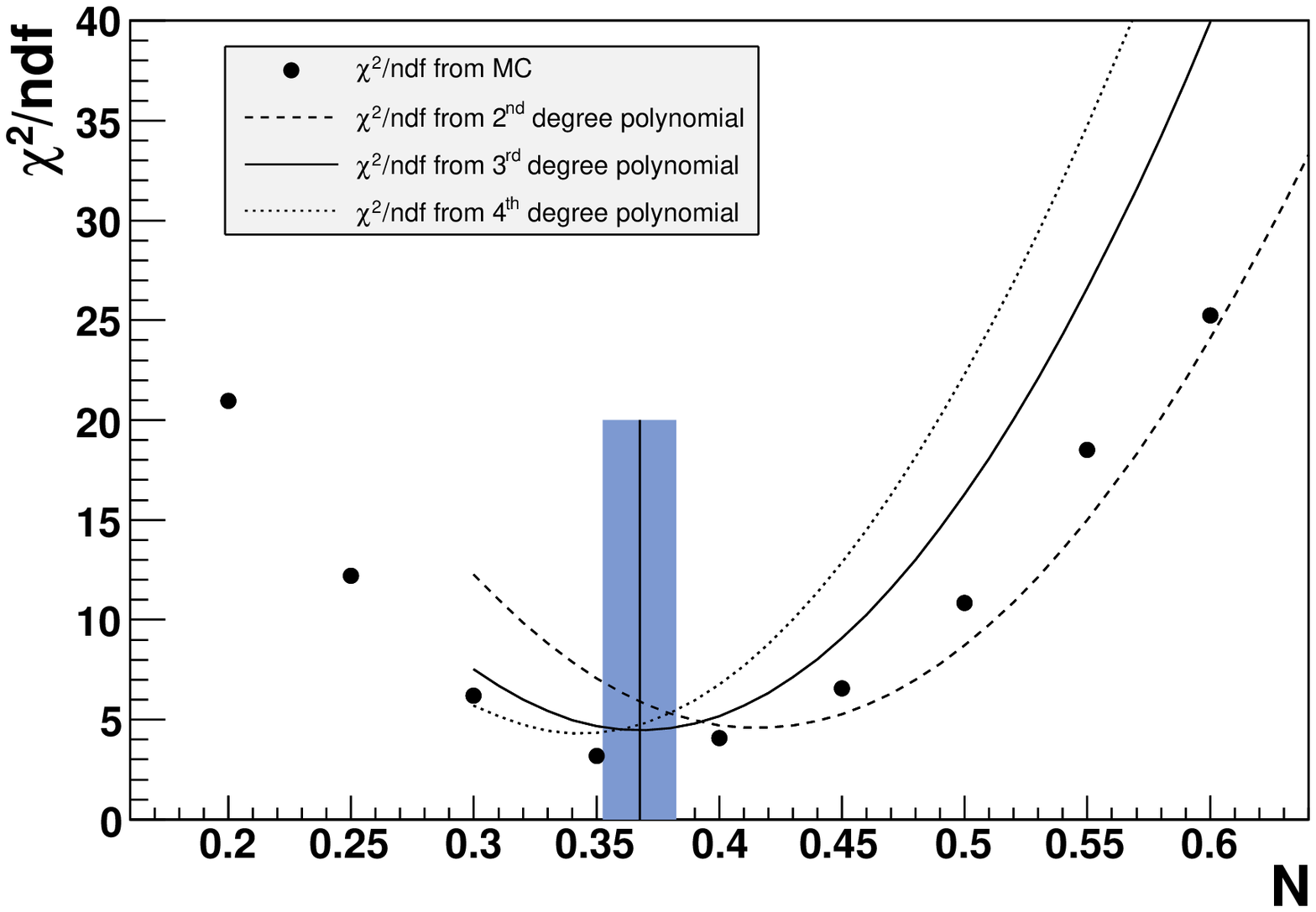}
\hfill
\includegraphics[width=0.33\columnwidth]{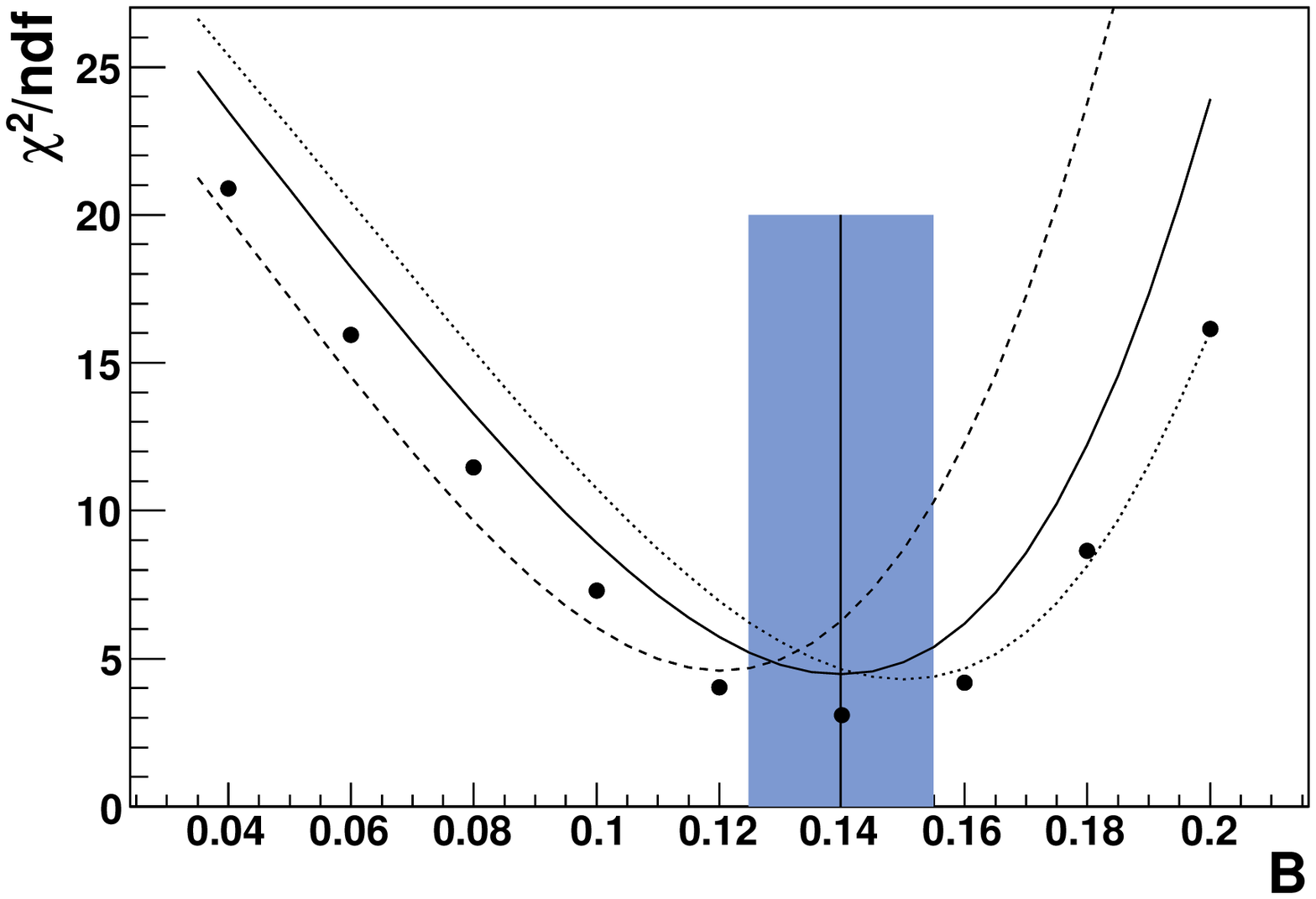}
\hfill
\includegraphics[width=0.33\columnwidth]{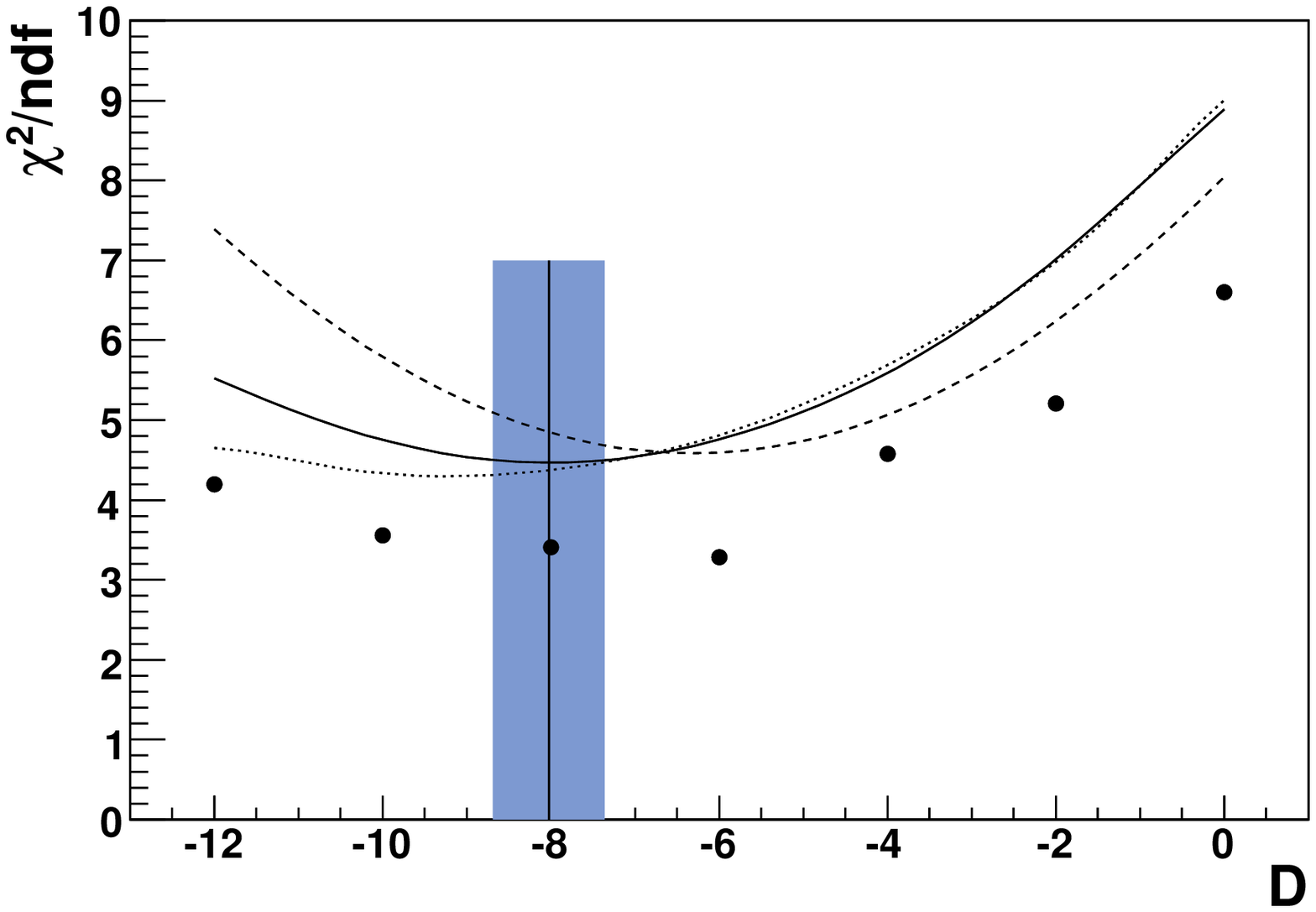}
}
\caption{Same as Fig.~\ref{f:chi2scans} but for Fit 4.}
\label{f:chi2scans4}
\end{figure}

In order to check if our minimization approach works, we 
scanned the parameter values
around their best values to check if we can indeed identify the signs
of the presence of a minimum of $\chi^2$. 
In Fig.~\ref{f:chi2scans} and \ref{f:chi2scans4} we show
how $\chi^2$
 changes as a function of each of the three parameters used in
Fit 2 and Fit 4, while fixing the other two to their best-fit value. 
The scans were carried out
using both the Monte Carlo generator directly and the grid approximation. For
comparison, we
show also the results obtained using second- and fourth-degree polynomial approximations.
First of all, we observe that 
the profile has in all cases a parabolic shape and the position of the minimum is
clearly visible. This gives us once again confidence in the reliability of the
procedure. We see also that the position of the minimum and the 
shape of $\chi^2$ from the MC computation are similar to what is obtained from
the grid approximation with third-degree polynomials. The position of the
minimum is similar to what is found using the fourth-degree polynomial
approximation, but quite different to what is found using the second-degree polynomial approximation.
The value of
the minimum $\chi^2$ is not the same for MC and grid approximation (1.4 versus
1.6 for Fit 2; 3.2 versus
4.6 for Fit 4). This is due to
the fact that the approximation errors, 
$\delta Y_i^{\rm app}$ in Eq.~\eqref{e:chi2exp}, are typically
smaller than the MC errors, $\delta Y_i^{\rm{MC}}$ in
Eq.~\eqref{e:chi2brute}, 
and lead to a higher $\chi^2$. This difference becomes irrelevant only if
$\delta Y_i^{\rm{MC}}$  is negligible compared to the experimental errors $\delta Y_i^{\rm{ex}}$.


%Of course, the fact that in this particular case the fitting procedure we used
%gives good results does not guarantee that the same will happen for
%different observables. Every time the results of the fit should be critically
%examined. We plan to apply this procedure to different observables in the future. 
%
%Once we have established that the results we obtained are reliable, 
At this point, we can briefly
discuss the physical meaning of our results.
First of all, we can conclude that in the
extended $x_{Bj}$ and $Q^2$ range of Fit 3 and 4 we cannot achieve a 
good description of the $F_2$ data with {\sc Cascade}. This is not surprising,
since the generator starts from a purely gluonic distribution function. The description
is in general better at lower values of $x_{Bj}$, where gluons dominate.

Secondly, we conclude that in the restricted $x_{Bj}$ and $Q^2$ range of Fit 1
and 2, a good description of the data is obtained when we include parameter $D$ to give more flexibility to the functional form of the
gluon distribution. 

A few considerations can be made also on the value of parameter $B$, governing
the low-$x$ behavior of the gluon distribution. In all fits, the value is
higher than previous studies~\cite{Hansson:2003xz,Jung:2004gs,Jung:2006ji}. 
This is for instance the reason of the different behaviors at
low $x_{Bj}$ and $Q^2$ in Fig.~\ref{f:case1}. The value of $B$ turns
out to be even higher in the fits with a free $D$ parameter.

Not surprisingly, 
we observe that the parameters of the gluon distribution function are in
general different from the ones obtained in global fits at similar input
scales~\cite{Gluck:2007ck,Martin:2009iq}. To start with, 
global fits include many more data
sets than we presently considered. However, there are more fundamental 
differences between the physics included in the generators or in global
fits. Therefore, to achieve the best possible description of data with Monte
Carlo generators, the parameters of the distribution functions should be tuned
independently of global fits.

%%%%%%%%%%%%%%%%%%%%%%%%%%%%%%%%%%%%%%%%%%%%%%%%%%%%%%%%%%%%%%%%%%%%%%%%%%%%%%%%%%%%%%%%%%%%%%%%%%%%%%%%%%%%%%%%%%%%%%%%%%%%%%%%%%%%%%%%%%%%%%%%%%%%%%%
\section{Conclusions}
%%%%%%%%%%%%%%%%%%%%%%%%%%%%%%%%%%%%%%%%%%%%%%%%%%%%%%%%%%%%%%%%%%%%%%%%%%%%%%%%%%%%%%%%%%%%%%%%%%%%%%%%%%%%%%%%%%%%%%%%%%%%%%%%%%%%%%%%%%%%%%%%%%%%%%%

In this work we analyzed a method to tune the parameters of Monte
Carlo event generators using a set of experimental observables.
First, the generator is run with a few
different values of the parameters to tune. For each observable, a grid of
predictions is thus obtained. The resulting 
grids are approximated by analytic functions of the parameters. Finally, 
the analytic functions are used in place of the generator itself to perform a
$\chi^2$ fit 
to the data and obtain the best values for the parameters. The method is
significantly faster than a direct use of the generator, as the construction of
the grids typically requires fewer calls to the generator than a direct fit
and all grid points can be computed in parallel. 
There is no need to rerun
the generator to repeat the fit with different initial values of the
parameters, nor if the experimental data change (for instance if the statistics
increase). If data for different observables become available, the generator
has to be run to build grids for these new observables, but the old grids
remain still valid for the old observables.

The main limit of the approach is that the
limited parameter ranges have to be fixed a priori, 
since the
grids have to be built once and for all before the fitting is actually
performed.
It is possible to improve the choice of the parameter ranges with
hindsight, after the first attempt. However, this approach might be time
consuming and the minimization can still
fail if the data cannot constrain the value of one or
more parameters.

As a concrete example, we applied the method to find the best values for the
parameters of the unintegrated gluon distribution function used in the
{\sc Cascade} Monte Carlo generator. To constrain the parameter values, 
we used the data on the $F_2$ structure function in inclusive deep inelastic
scattering.
%measured by the H1
%Collaboration~\cite{Adloff:2000qk}.  
We performed four different types of fit, changing the range of $x_{Bj}$ and
$Q^2$ and the number of free parameters under consideration. 

Taking the second version of the fit as an illustration, we chose 150 combinations of parameter
values and produced a grid of predictions for each one of the 122 data points. 
The
grid was approximated by a third order polynomial with a total of
35 coefficients. The best approximation was searched for using the method of
Single Value Decomposition to guarantee a fast and reliable search. The
quality of the approximation was found to be very good, with $\chi^2/{\rm
  n.d.f.}=1.05$.

Finally, we found the best values of the parameters by a second
 $\chi^2$ minimization, using the
difference between the experimental measurements and the analytic
approximation of the generator output to define the $\chi^2$
function. The minimization was done using MINUIT. 

%Once the best-fit values of the parameters are found, we compute the final
%prediction using {\sc Cascade} directly, not its analytical approximation. 
We
checked that the best-fit values of the parameters give a good 
description of the data, with a $\chi^2/{\rm
  d.o.f.}=1.6$. 
By scanning
the dependence of $\chi^2$ on the single parameters, we strengthened the
evidence that the fit found the parameter values that describe the data best.  

By including more data in the fit, the method described in this work 
can be applied to better constrain the parameters of the unintegrated gluon
distribution, including those describing the
intrinsic $k_t$-dependence.
%Further studies are required to improve the fitting procedure and to
%better fix the parameters of the unintegrated gluon distribution function, for
%instance by
%including data sets that are more sensitive to the details
%of its $k_t$-dependence (e.g., dijet-production data~\cite{Aktas:2003ja}).

%%%%%%%%%%%%%%%%%%%%%%%%%%%%%%%%%%%%%%%%%%%%%%%%%%%%%%%%%%%%%%%%%%%%%%%%%%%%%%%%%%%%%%%%%%%%%%%%%%%%%%%%%%%%%%%%%%%%%%%%%%%%%%%%%%%%%%%%%%%%%%%%%%%%%%%
\acknowledgments
%%%%%%%%%%%%%%%%%%%%%%%%%%%%%%%%%%%%%%%%%%%%%%%%%%%%%%%%%%%%%%%%%%%%%%%%%%%%%%%%%%%%%%%%%%%%%%%%%%%%%%%%%%%%%%%%%%%%%%%%%%%%%%%%%%%%%%%%%%%%%%%%%%%%%%%
Valuable discussions with Hendrik d'Hoeth are thankfully acknowledged. 
The work of A.B. was partially supported by the SFB ``Particles, Strings and the
Early Universe'' and partially by U.S.\ DOE Contract No.~DE-AC05-06OR23177
under which JSA operates Jefferson Laboratory.

%%%%%%%%%%%%%%%%%%%%%%%%%%%%%%%%%%%%%%%%%%%%%%%%%%%%%%%%%%%%%%%%%%%%%%%%%%%%%%%%%%%%%%%%%%%%%%%%%%%%%%%%%%%%%%%%%%%%%%%%%%%%%%%%%%%%%%%%%%%%%%%%%%%%%%%

\appendix

\section{Treatment of correlated systematic uncertainties} 
\label{treatmentofcorrelatedsystematicerrors} 
A convenient
method to determine the quality of a fit is to use a least square
minimization. This ansatz is justified by the assumption that the errors are
Gaussian distributed.  

A set of measurements \{$d_i$\} will in general
deviate from a set of corresponding predictions \{$t_i$\}. The deviations are
caused by various kinds of uncertainties as there is for each data point a
statistical uncertainty $\sigma_i^{{\rm dat}}$, an uncorrelated systematic
uncertainty $u_i$ and, coming from $n_{{\rm sys}}$ sources, the correlated
systematic  uncertainties $\{\beta_{i1},\beta_{i2},...,\beta_{in_{{\rm sys}}}\}$. 
The measurement is then related to the prediction by: 
\begin{equation} 
\label{datapointCTEQ} 
d_i=t_i+r_i\alpha_i+\sum_{k=1}^{n_{{\rm sys}}}r'_k\beta_{ik},
\end{equation} 
where $\sigma_i^{{\rm dat}}$ and $u_i$ are added in quadrature to form a unified
uncorrelated error $\alpha_i=\sqrt{{(\sigma_i^{{\rm dat}})}^2+{(u_i)}^2}$. The
$r_i$, $r'_k$ express the individual shifts of the data points by the
uncertainties and are Gaussian distributed with zero mean and unit variance
and assumed to be independent of each other.

A $\chi^2$ that includes a proper
treatment of correlated systematic errors can be calculated as follows (see
\cite{Stump:2001gu} for a derivation): 
\begin{equation}
\label{chi2CTEQ} 
\chi^2(\{a\},\{r'\})=\sum_{i=1}^N
\biggl(\frac{d_i-t_i-\sum_{k=1}^{n_{{\rm sys}}}\beta_{ik}r'_k)}{\alpha_i}\biggr)^2
+\sum_{k=1}^{n_{{\rm sys}}} {r'_k}^2 ,
\end{equation} 
where it can be seen that $\chi^2$ depends both on $\{a\}$ (the parameters
entering the predictions $t_i$) and the random parameters
$\{r'\}$. 
The latter ones can be expressed as 
\begin{equation} 
r'_k(\{a\})=\sum^{n_{{\rm sys}}}_{k'=1}(A^{-1})_{kk'}B_{k'}, 
\end{equation} 
which leads to the $r'$-independent form 
\begin{equation} 
\chi^2(\{a\})=\sum^{N_{{\rm dat}}}_{i=1}
\frac{(d_i-t_i)^2}{\alpha_i^2}-\sum^{n_{{\rm sys}}}_{k,k'=1}B_k~(A^{-1})_{kk'} B_{k'} 
\end{equation} 
with 
\begin{align} 
\label{ABcteq}
B_k&=\sum_{i=1}^{N_{{\rm dat}}}\frac{\beta_{ik}(d_i-t_i)}{\alpha_i^2}
&&{\rm and}  
&A_{kk'}&=\delta_{kk'}+
\sum_{i=1}^{N_{{\rm dat}}}\frac{\beta_{ik}\beta_{ik'}}{\alpha_i^2}
\end{align}
For the systematic errors, in this work we used the ansatz
proposed by CTEQ, i.e.,
\begin{equation} 
\beta_{ik}=\epsilon_{ik}d_i
\end{equation} 
with $\epsilon_{ik}$ being the relative systematic error.

\bibliographystyle{myrevtex}
\bibliography{updfitting_biblio}

\end{document}